\newcommand{\ra}{\rightarrow}
\newcommand{\vub}{|V_{ub}|}
\newcommand{\vcb}{|V_{cb}|}
\newcommand{\btodslnu}{\bar{B}\ra D^* \ell\bar{\nu}}
\newcommand{\btopilnu}{\bar{B}\ra \pi\ell\bar{\nu}}
\newcommand{\lone}{\lambda_1}
\newcommand{\ltwo}{\lambda_2}
\newcommand{\Lbar}{\bar{\Lambda}}
\newcommand{\btoclnu}{b\ra c\ell\bar{\nu}}
\newcommand{\btoulnu}{b\ra u\ell\bar{\nu}}
\newcommand{\btosgamma}{b\ra s\gamma}

\documentclass[epj]{svjour}
\usepackage{graphics}
\usepackage{times}
\begin{document}
\title{$|V_{cb}|$, $|V_{ub}|$, and HQET at CLEO}
\author{Karl M.~Ecklund\inst{1}}
\institute{Floyd R.~Newman Laboratory for Elementary-Particle Physics,
Cornell University, Ithaca, New York, 14853}
\date{Received: November 1, 2003}
\abstract{
I report results from the CLEO collaboration on semileptonic $B$
decays, highlighting measurements of the Cabibbo-Kobayashi-Maskawa
matrix elements $\vcb$ and $\vub$. I describe the techniques used to
obtain the recent improvements in precision for these measurements,
including the use of the $b\ra s\gamma$ photon spectrum to reduce
hadronic uncertainties in semileptonic $B$ decays.  I also report new
measurements of $\vcb$ using the inclusive semileptonic branching
fraction  ${\cal B}(B\to X e \nu)$ and of $\vub$ through study of the
$q^2$ dependence of $B\to\pi\ell\nu$ and $B\to\rho\ell\nu$.
\PACS{
      {12.15.Hh}{Determination of Kobayashi-Maskawa matrix elements}   \and
      {13.20.He}{Decays of bottom mesons}
     } 
} 
\maketitle
\section{Introduction}
\label{intro}
The study of semileptonic $B$ meson decays allows measurement of the
Cabibbo-Kobayashi-Maskawa (CKM) matrix elements $\vcb$ and $\vub$,
providing important inputs to a test of the unitarity of the CKM
matrix, which governs the weak charged current and gives rise to $CP$
violation in the standard model.
The rate for a $b$ hadron to decay weakly to hadrons containing a $c$
or $u$ quark is proportional to $\vcb^2$ or $\vub^2$ respectively.
The absence of final-state interactions in semileptonic decay make the
interpretation less dependent on hadronic matrix elements than fully
hadronic $B$ decays, although hadronic uncertainties still limit the
precision of $\vub$ and $\vcb$ measurements.

The current round of measurements from CLEO continues to test the
hadronic calculations needed to disentangle weak matrix elements
from strong interaction effects.
For decays of $B$ mesons to exclusive final states, the hadronic
effects are expressed in terms of a form factor that depends only on
the momentum transfer $q^2$ to the lepton neutrino pair.  By measuring
decay rates as a function of $q^2$ we have begun to test the form
factors, particularly for $\btoulnu$ transitions.
In decays to inclusive final states, under the assumption of
parton-hadron duality, quark-level calculations may be compared to
inclusive measurements to extract CKM matrix elements.  Measurement of
spectral distributions in inclusive decays gives additional
observables to overconstrain theory parameters and test how well the
theory and parton-hadron duality works.

\section{$\vcb$ Measurements}
\label{sec:vcb}
\begin{figure}
\begin{center}
\resizebox{7cm}{!}{\includegraphics{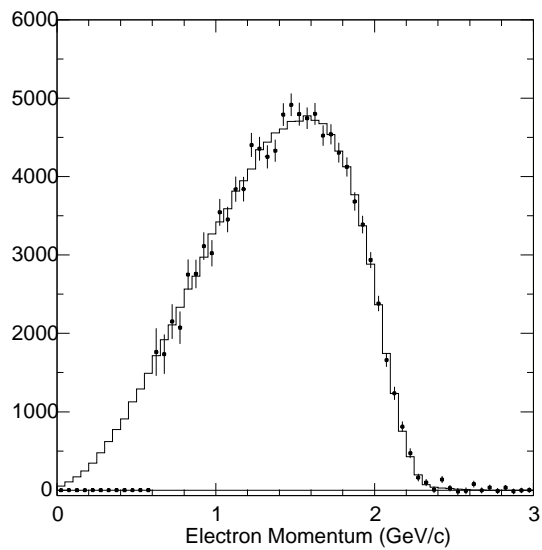}}
\end{center}
\caption{Unfolded primary $B \to X e^+\nu$ electron spectrum measured
using a high momentum lepton tag. The line shows a fit to exclusive
$B\to X e^+\nu$ decays, used to extrapolate below the cut at 600
MeV/$c$. (Preliminary)}
\label{kme:fig.b2xenu}
\end{figure}

A measurement of $\vcb$ is possible using the inclusive semileptonic
decay rate.  The experimental inputs are the branching 
fraction for $\bar{B}\to X_c\ell\bar{\nu}$ and the $B$ lifetime.  The
inclusive decay rate $\Gamma_c^{SL} = \gamma_c \vcb^2$, where
$\gamma_c$ comes from theory.
Within the framework of heavy quark effective theory (HQET)
\cite{ManoharWise:2000dt}, the inclusive semileptonic
decay rate is expanded in a double series in $\alpha_s^n$ and
$1/M^n$, where $M$ is the heavy quark mass.  Hadronic effects enter
both in the perturbative expansion and as expansion parameters,
defined to be matrix elements of non-perturbative QCD operators.  At
${\cal O}(1/M^2)$ there are two parameters: $\lone$, which is
proportional to the kinetic energy of the $b$ quark in the $B$ meson,
and $\ltwo$, which comes from the chromomagnetic operator.  An
additional parameter $\Lbar$ relates the $B$ meson mass to the $b$
quark mass. From the $B$-$B^*$ mass difference, $\ltwo=0.128\pm0.010$
GeV$^2$. The other parameters can be estimated (\textit{e.g.}~in quark
models) but they can also be measured using spectral moments in
inclusive $B$ decay. Moments, \textit{e.g.}~of the lepton energy
spectrum, are also computed in HQET, allowing extraction of $\lone$
and $\Lbar$ from two or more spectral measurements.

\begin{figure*}
\resizebox{!}{6cm}{\includegraphics{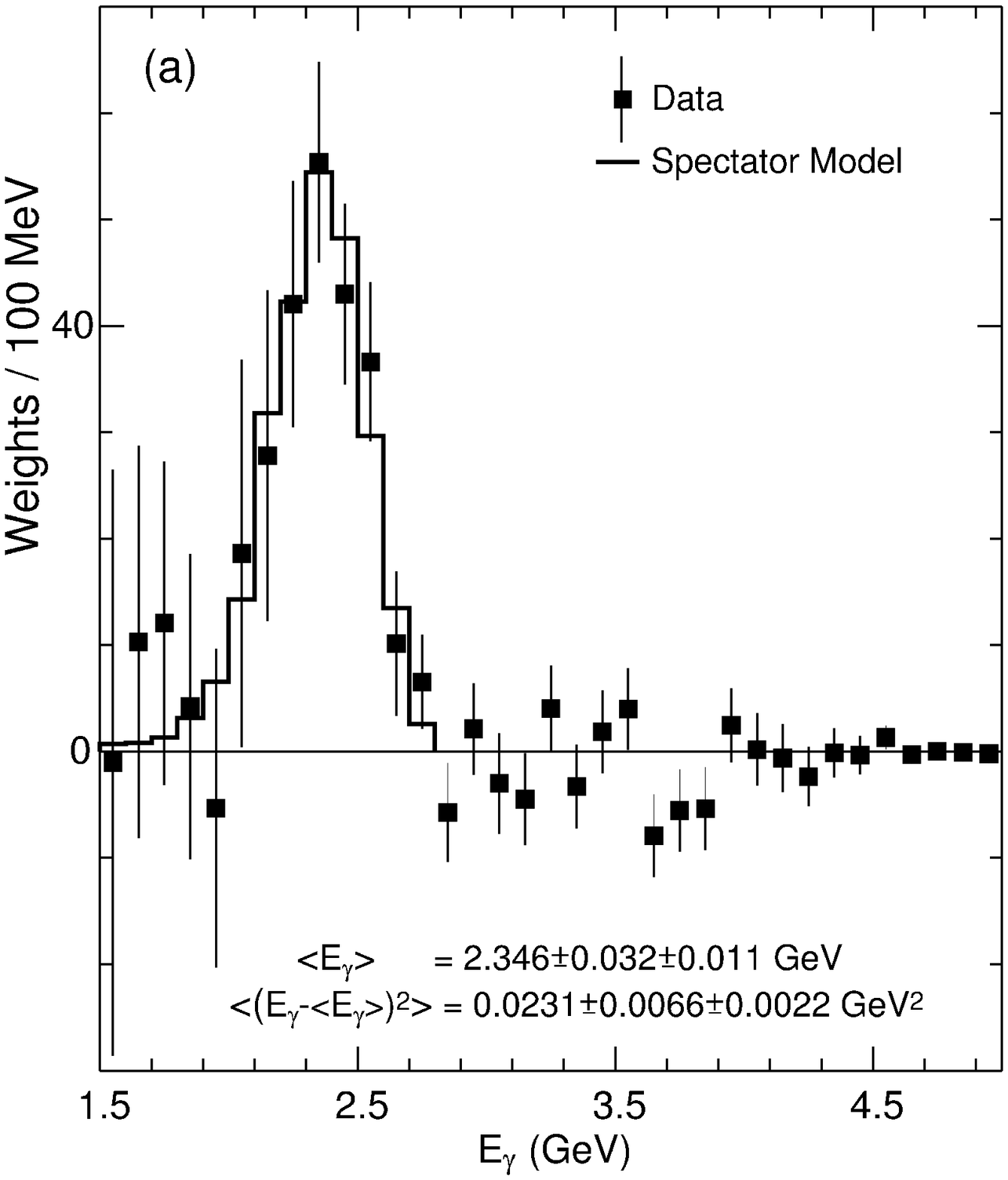}}
\resizebox{!}{6cm}{\includegraphics{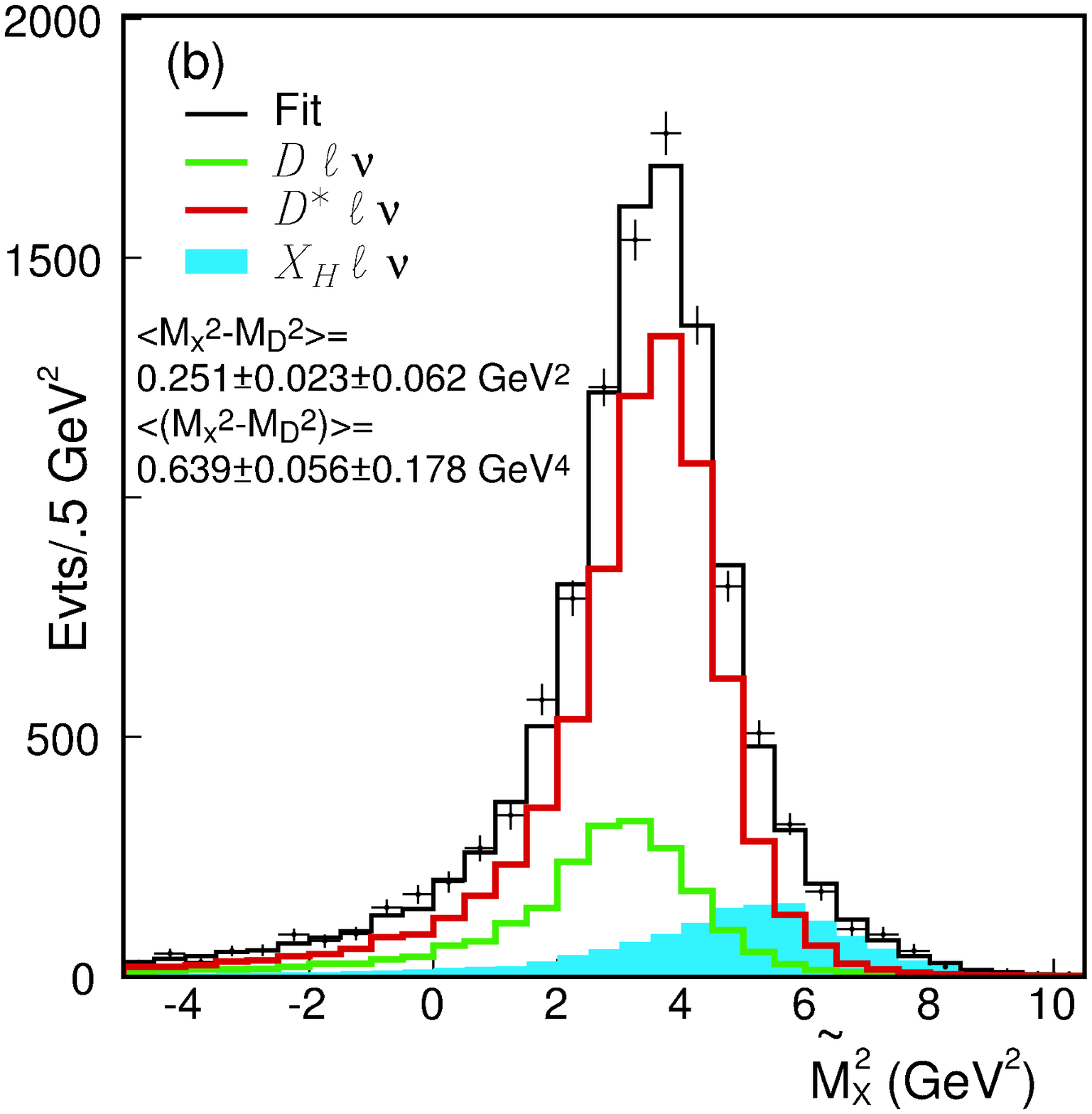}}
\resizebox{!}{6cm}{\includegraphics*{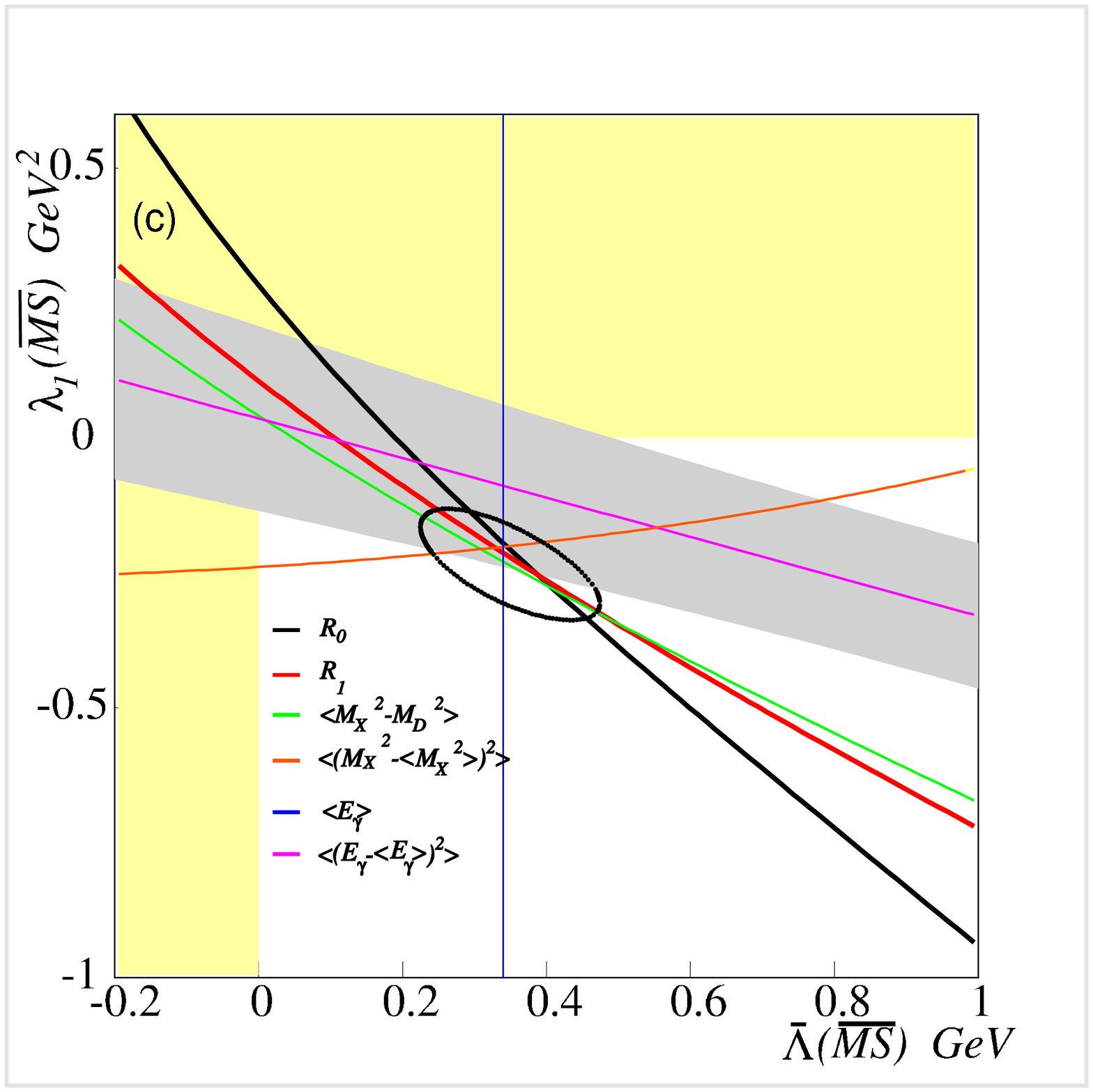}}
\caption{$B\to X_s\gamma$ photon spectrum (a), $\bar{B}\to X_c\ell\bar{\nu}$ 
$M_X^2$ spectrum for $p_\ell>1.5$ GeV/$c$ (b), and constraints on HQET
parameters (c) from CLEO moment measurements.  As an illustration of
theoretical uncertainties, the shaded band includes ${\cal O}(1/M^3)$
uncertainties for $\langle (E_\gamma -\langle E_\gamma \rangle)^2 \rangle$.}  
\label{kme:fig.hqet}
\end{figure*}

CLEO has a new preliminary measurement of the inclusive semileptonic
branching fraction using a high-momentum ($p>1.5$ GeV/$c$) lepton tag.
The analysis is an update of Ref.~\cite{Barish:1996cx}, where the
lepton tag  identifies a sample of $B$ decays with high purity (98\%).
Additional electrons may come from the decay chain of the same $B$ or
from the decay of the other $B$ meson in the event 
($e^+e^-\ra\Upsilon(4S)\ra B\bar{B}$). 
Secondary electrons ($b\to c\to e$) and primary electrons are
separated using kinematic and charge correlations, with a known
correction from $B^0$-$\bar{B^0}$ mixing.  The new semileptonic
branching fraction is  $10.88\pm0.08\pm0.33$\%.  The spectrum of
electrons above 600 MeV/$c$ is also obtained
(Fig.~\ref{kme:fig.b2xenu}), from which spectral moments will be measured.

CLEO has recently measured spectral moments in inclusive semileptonic
decay and in $B\to X_s\gamma$.  These are used to extract HQET
parameters and reduce the theoretical uncertainty in inclusive $\vcb$
measurements.
CLEO measured the $B\to X_s\gamma$ photon spectrum and moments
(Fig.~\ref{kme:fig.hqet}a) in \cite{Chen:2001fj}.  
In \cite{Cronin-Hennessy:2001fk}, CLEO measured the moments of the
hadronic mass distribution in $\bar{B}\to X_c\ell\bar{\nu}$ decays
with $p_\ell > 1.5 $ GeV/$c$ (Fig.~\ref{kme:fig.hqet}b).
Combining the constraints on $\lone$ and $\Lbar$ from the first
moments of the photon energy and hadronic mass spectra, we obtain a solution
for $\lone$ and $\Lbar$ and extract 
$\vcb = (41.1 \pm 0.5_{\lone,\Lbar} 
              \pm 0.7_{\Gamma}
              \pm 0.8_{HQET}) \times 10^{-3}$
using the new CLEO branching fraction and PDG2003 lifetime average
as inputs.  The uncertainties from unknown ${\cal O}(1/M^3)$ HQET
parameters are dominant.

The lepton energy moments in $\bar{B}\to X_c\ell\bar{\nu}$ are also
sensitive to the HQET parameters, and CLEO has measured the lepton
spectrum \cite{Mahmood:2002tt} and normalized moments $R$
\cite{Gremm:1996yn} above 1.5 GeV. From all of the moment
measurements, one can assemble the constraints on the HQET parameters
$\lone$ and $\Lbar$. Figure~\ref{kme:fig.hqet}c shows the remarkable
consistency of these measurements, lending credibility to the
inclusive $\vcb$ measurement. 

At present the inclusive $\vcb$ measurement is more precise (3\%) than
that from $\btodslnu$, but with reliance on HQET for hadronic
corrections.  The first tests of HQET using spectral moments in
inclusive $B$ decays give us some confidence in the method, but
additional tests with more inclusive moments are needed.  This summer
BaBar \cite{langenegger} and CLEO \cite{lipeles} presented a new
analyses of the hadronic mass spectral moments with lepton energy as
low as 0.9 GeV, in good agreement with previous measurements and
expectations from HQET within the limits of ${\cal O}(1/M^3)$
uncertainties.

The agreement between inclusive and exclusive measurements is another
test of our control of hadronic corrections.  There is good agreement
between inclusive and the world average exclusive $\vcb$ measurements
\cite{stone:eps03}, but the confidence level of the $\btodslnu$ world
average is presently poor \cite{HFAG:sum03}.

\section{$\vub$ Measurements}
\label{sec:vub}
Measurements of $\btoulnu$ have to contend with a 50--100 times larger
background from $\btoclnu$.  Requiring a lepton energy above
the endpoint for $\btoclnu$ ($\approx 2.3$ GeV) is the easiest
strategy to reduce background, but this cut near the edge of the
spectrum introduces sensitivity to the motion of $b$ quark in the $B$
meson.  The sensitivity is reduced by using the $\btosgamma$ photon
spectrum \cite{Chen:2001fj}, which is sensitive to the same hadronic
effects at leading order 
\cite{Neubert:1994ch,Neubert:1994um,Bigi:1994ex,Leibovich:1999xf}.

CLEO measured the lepton spectrum from $B$ decays in the endpoint
region $E>2.2$ GeV and extracted a partial branching fraction of
$(2.30 \pm 0.15 \pm 0.35) \times 10^{-4}$ \cite{Bornheim:2002du}.
From the $\btosgamma$ photon spectrum, the fraction of $\btoulnu$
events passing the lepton energy cut is $f_u = 0.130 \pm 0.024 \pm 0.015$.
This gives 
$|V_{ub}| = (4.08 \pm 0.34_{\rm exp} \pm 0.44_{f_u} \pm 0.16_{\Gamma}
\pm 0.24_{NLO}) \times 10^{-3}$, where the theoretical
uncertainties are $\Gamma$, from \cite{Hoang:1998hm,Uraltsev:1999rr},
and $NLO$, from sub-leading terms relating hadronic effects in
$\btoulnu$ and $\btosgamma$.

\begin{figure*}
\resizebox{!}{6cm}{\includegraphics{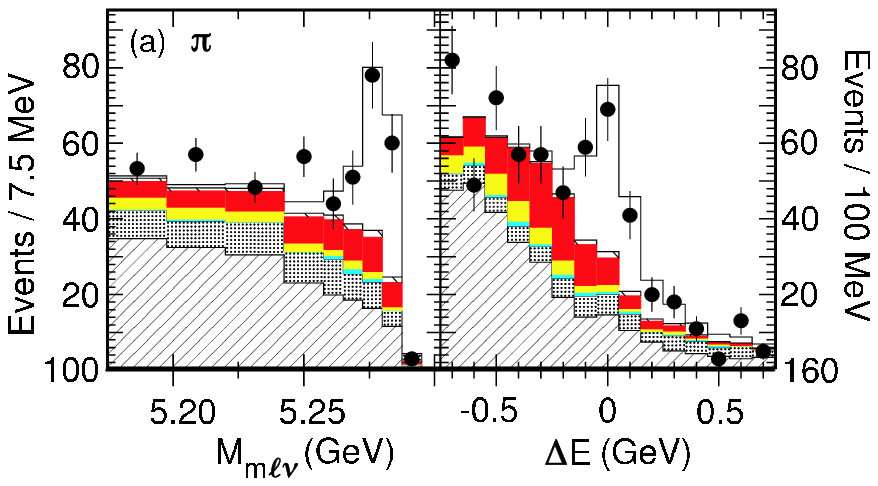}}
\resizebox{!}{6cm}{\includegraphics{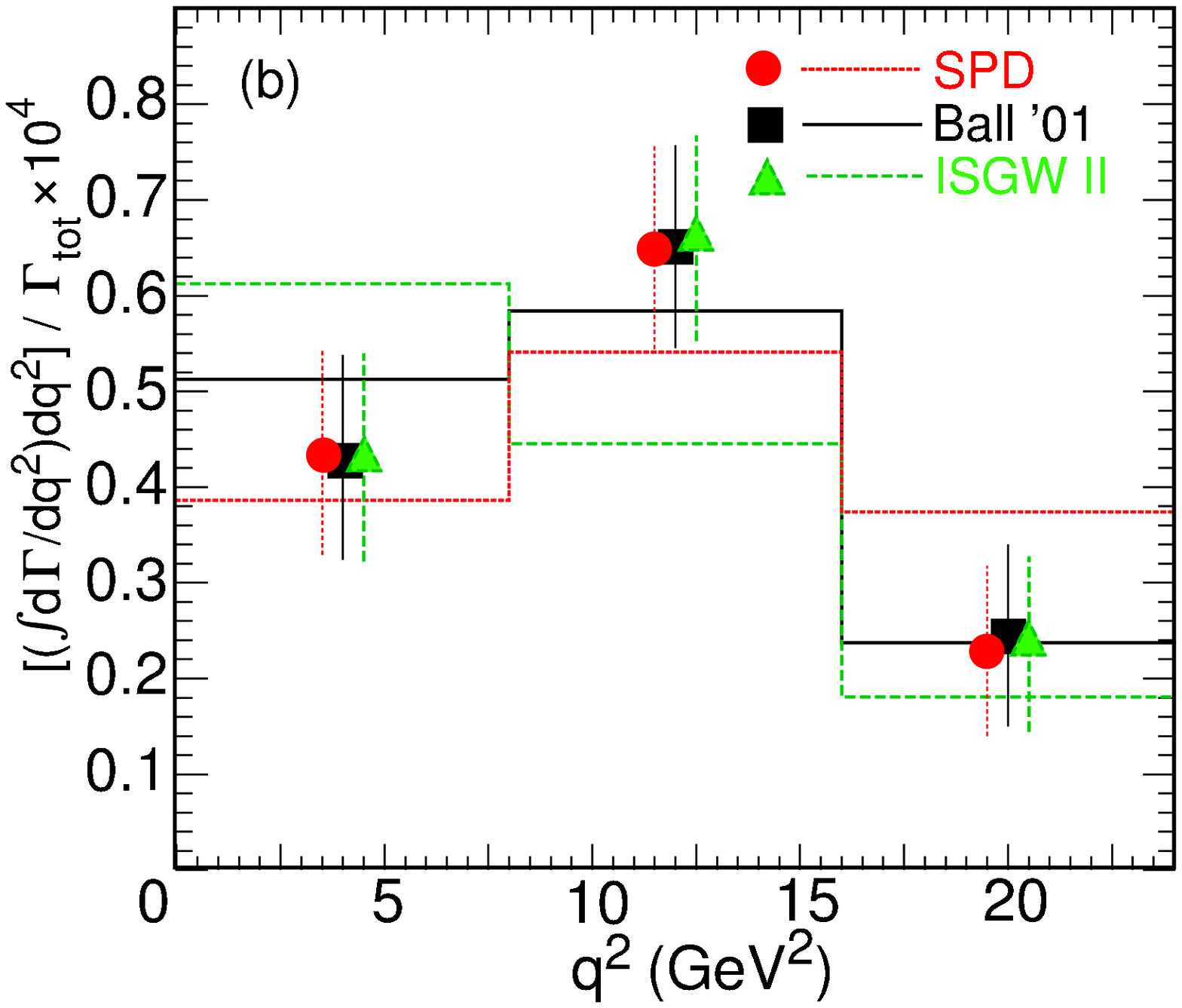}}
\caption{Exclusive $\btopilnu$: (a) projections of ML fit to
$M_{m\ell\nu}$ and $\Delta E$ and (b) fit to $d\Gamma/dq^2$.}
\label{kme:fig.ExcVub}
\end{figure*}

In a contribution to this conference, CLEO has also measured $\vub$ in
the exclusive modes ${\bar B}\to [\pi/\rho/\omega/\eta]\ell\bar{\nu}$
\cite{Athar:2003yg},  where kinematic constraints from the full 
reconstruction of the final state gives the needed suppression of
$\btoclnu$.  The neutrino is reconstructed from the missing energy and
momentum of the event, taking advantage of CLEO's large solid angle
(95\%).  Combined with a lepton and light meson candidate, energy and
momentum conservation leads to signal peaks in $\Delta E = E - E_{\rm
beam}$ and the $B$ candidate invariant mass $M_{m\ell\nu}$, with $S/B
\approx 1$.  We perform a simultaneous maximum likelihood fit in
$\Delta E$ and $M_{m\ell\nu}$ to seven sub-modes: 
$\pi^\pm$, $\pi^0$, $\rho^\pm$, $\rho^0$, 
$\omega/\eta \to \pi^+\pi^-\pi^0$, and $\eta \to \gamma\gamma$.  
In the fit we use isospin symmetry to constrain the semileptonic widths
$\Gamma^{SL}({\pi^\pm})=2\Gamma^{SL}({\pi^0})$ and 
$\Gamma^{SL}({\rho^\pm})=2\Gamma^{SL}({\rho^0})\approx 2\Gamma^{SL}({\omega})$,
where the final approximate equality is inspired by constituent quark
symmetry.
Signals for $\pi$ (Fig.~\ref{kme:fig.ExcVub}a) and $\rho$ are extracted
separately in three $q^2$ bins.  Given form factors from theory, we
extract $\vub$ from a fit to $d\Gamma/dq^2$ (Fig.~\ref{kme:fig.ExcVub}b). 
Combining $\btopilnu$ and
$\bar{B}\to\rho\ell\bar{\nu}$ results we find 
\begin{displaymath}
|V_{ub}| =  (3.17 \pm 0.17           |_{\rm stat}
                    \ ^{+0.16}_{-0.17} |_{\rm syst}
                    \ ^{+0.53}_{-0.39} |_{\rm theo}
                    \pm 0.03           |_{\rm FF}   ) \times 10^{-3}.
\end{displaymath}
This result uses form factors from Lattice QCD ($q^2 > 16$ GeV$^2$) and
light cone sum rules ($q^2 > 16$ GeV$^2$) where each are most reliable.
In a test of $\btopilnu$ form factors, ISGW2
\cite{Scora:1995ty} is disfavored (Fig.~\ref{kme:fig.ExcVub}b).

We find good agreement between measurements of $\vub$ using inclusive
and exclusive techniques.  The theoretical uncertainty on the form
factor normalization currently limits the precision of the exclusive
$\vub$ measurement.  In the future, unquenched Lattice QCD
calculations can improve the $\btopilnu$ form factor in a limited
region of $q^2$. Inclusive $\btoulnu$ measurements can be 
further improved with increased $\btosgamma$ statistics and better
phenomenological understanding of non-perturbative shape functions
for the $B$ meson \cite{Leibovich:2002ys,Bauer:2002yu,Neubert:2002yx}.  
Comparison between inclusive measurements that use
kinematic cuts that are less dependent on hadronic effects
(more inclusive and away from the endpoint region) will increase our
confidence in inclusive $\vub$ measurements.  Since the principal
background comes from $\btoclnu$, better knowledge of the dominant
semileptonic $B$ decays will improve systematic errors for both
inclusive and exclusive measurements.

\end{document}